\documentclass[11pt]{article}
\usepackage{amssymb}

\newcommand{\Hajek}{H\'{a}jek}

\newcommand{\Pudlak}{Pudl\'{a}k}

\newcommand{\Nepom}{Nepomnja\v{s}\v{c}i\v{\i}}
\newcommand{\Zak}{\v{Z}\`{a}k}

\newtheorem{thm}{Theorem}

\newtheorem{cor}{Corollary}

\newtheorem{lem}{Lemma}

\newtheorem{prop}{Proposition}

\newcommand{\compfont}{\mathsf}

\newcommand{\UNIV}{\compfont U}
\newcommand{\EXIST}{\compfont E}
\newcommand{\SEXIST}{\compfont e}
\newcommand{\SUNIV}{\compfont u}

\newcommand{\lin}{\compfont{lin}}

\newcommand{\NP}{\compfont{NP}}
\newcommand{\coNP}{\compfont{co}$-$\compfont{NP}}
\newcommand{\PH}{\compfont{PH}}
\newcommand{\BH}{\compfont{BH}}
\newcommand{\LINH}{\compfont{LinH}}
\newcommand{\EuLINH}{\compfont{Eu}\mbox{-}\compfont{LinH}}
\newcommand{\UeLINH}{\compfont{Ue}\mbox{-}\compfont{LinH}}

\newcommand{\xiLINH}{\xi\mbox{-}\compfont{LinH}}

\newcommand{\SC}{\compfont{SC}}

\newcommand{\co}{\compfont{co}}
\newcommand{\DLIN}{\compfont{DLIN}}

\newcommand{\NLIN}{\compfont{NLIN}}
\newcommand{\coNLIN}{\compfont{co}$-$\compfont{NLIN}}
\newcommand{\LOGSPACE}{\compfont{L}}

\newcommand{\NLOGSPACE}{\compfont{NL}}
\newcommand{\coNLOGSPACE}{\compfont{co}$-$\compfont{NL}}
\newcommand{\DTIME}{\compfont{DTIME}}
\newcommand{\TIME}{\compfont{TIME}}
\newcommand{\NTIME}{\compfont{NTIME}}
\newcommand{\DSPACE}{\compfont{DSPACE}}

\newcommand{\NSPACE}{\compfont{NSPACE}}
\newcommand{\PTIME}{\compfont{P}}
\newcommand{\TISP}{\compfont{TISP}}
\newcommand{\DTISP}{\compfont{DTISP}}
\newcommand{\NTISP}{\compfont{NTISP}}
\newcommand{\SigmaP}[1]{\Sigma^{\compfont p}_{#1}}
\newcommand{\PiP}[1]{\Pi^{\compfont p}_{#1}}
\newcommand{\poly}{\compfont{poly}}
\newcommand{\pad}{\compfont{pad}}

\newcommand{\monus}{\mathbin{\mathchoice%
{\buildrel .\lower.6ex\hbox{\vphantom{.}} \over {\smash-}}%
{\buildrel .\lower.6ex\hbox{\vphantom{.}} \over {\smash-}}%
{\buildrel .\lower.4ex\hbox{\vphantom{.}} \over {\smash-}}%
{\buildrel .\lower.3ex\hbox{\vphantom{.}} \over {\smash-}}}}

\newenvironment{proof}%
{
\noindent {\it Proof.}
}{$\Box$}

\newcommand{\ignore}[1]{}

\begin{document}

\date{{\em \today}}
\title{Alternating Hierarchies for Time-Space Tradeoffs }
\author{Chris Pollett\\
Department of Computer Science\\
San Jose State University \\
1 Washington Square\\
San Jose, CA 95192\\
pollett@cs.sjsu.edu
\and
Eric Miles\\
Department of Computer Science\\
San Jose State University\\
1 Washington Square\\
San Jose, CA 95192\\
enmiles@gmail.com
}
\maketitle

\thispagestyle{empty}

\begin{abstract}
\Nepom's Theorem states that for all $0 \leq \epsilon < 1$ and $k>0$  the class of languages recognized in nondeterministic time $n^k$ and space $n^\epsilon$, $\NTISP[n^k, n^\epsilon]$, is contained in the linear time hierarchy. By considering restrictions on the size of the universal quantifiers in the linear time hierarchy, this paper refines \Nepom's result to give a sub-hierarchy, $\EuLINH$, of the linear time hierarchy that is contained in $\NP$ and which contains $\NTISP[n^k, n^\epsilon]$. Hence, $\EuLINH$ contains $\NLOGSPACE$ and $\SC$.  This paper investigates basic structural properties of $\EuLINH$. Then the relationships between $\EuLINH$ and the classes $\NLOGSPACE$, $\SC$, and $\NP$ are considered to see if they can shed light on the $\NLOGSPACE = \NP$ or $\SC=\NP$ questions.  Finally, a new hierarchy, $\xiLINH$, is defined to reduce the space requirements needed for the upper bound on $\EuLINH$.


\end{abstract}

\section{Introduction}

Fortnow~\cite{fortnow00}  began the study of time-space trade-offs as an approach to showing $\LOGSPACE \neq \NP$. His paper gives the first non-trivial time-space trade-offs for satisfiability.
These results were subsequently improved upon and the interested reader can consult Fortnow et al\@.~\cite{flmv06} and Williams~\cite{williams2005} for an introduction to the literature as well as current results. One tool used in these time-space results is \Nepom's Theorem. The form of this theorem we are interested in shows that $\NTISP(n^k, n^\epsilon)$, the class of languages decidable simultaneously in nondeterministic time $n^k$ and space $n^\epsilon$ for 
 $0 < \epsilon < 1$ and $k>0$, is contained in the linear time hierarchy, $\LINH$. If one examines the proof of this result, one is struck that although a language $L$ in $\NTISP(n^k, n^\epsilon)$ is seemingly mapped into the $\Sigma^{\lin}_{2k+1}$ level of the linear-time hierarchy, the universal quantifiers that appear are all of logarithmic size. It is thus natural to ask what is the complexity of the sub-class of  $\Sigma^{\lin}_{k}$ where we restrict the universal quantifiers to be logarithmic size. It is this hierarchy that we investigate in the present paper.

Given a class of languages  $\mathcal{C}$, let $\EXIST\mathcal{C}$ (resp\@., $\UNIV\mathcal{C}$) denote the class of languages of the form $\{ x \, | \, \exists y \in \{0,1\}^{k \cdot |x|}$, $\langle x, y \rangle \in L\}$ (resp\@., $\{ x \, | \, \forall y \in \{0,1\}^{k\cdot |x|}$, $\langle x, y \rangle \in L\}$) for some $L \in \mathcal{C}$ and fixed integer $k$.  (Note that we say $\{0,1\}^n$ to refer to the set of all binary strings with length no greater than $n$.)  We use lower case, $\SEXIST$ or $\SUNIV$ to denote the case where the bounding term $k\cdot |x|$ is replaced with  $\log (k\cdot |x|)$ for some fixed integer $k$. We will call these kind of quantifiers {\em sharply bounded}. Using this notation, we write $(\EXIST\SUNIV)^{\lin}_k$ for the class languages given by $k$ quantifier blocks whose values are finally fed into a deterministic linear time languages, where each quantifier block is of the form an $\EXIST$ quantifier followed by an $\SUNIV$ quantifier.   By examining the proof of \Nepom's theorem, we show every language in  $\NTISP[n^k, n^\epsilon]$  can be represented as a language in $\EuLINH := \cup_k (\EXIST\SUNIV)^{\lin}_k$. We in fact show a level-wise result with an almost matching upper bound: $\NTISP(n^{k(1-\epsilon)}, n^\epsilon)\subseteq(\EXIST\SUNIV)^{\lin}_{k}\subseteq  \NTISP[n^{k+1}, n] $. So our hierarchy is closely associated with the class of languages that can be solved using nondeterministic polynomial time and linear space.  As it is simultaneously contained in the linear time hierarchy and in $\NP$, it also contains $\NLOGSPACE$ and $\SC: = \cup_k\cup_m \DTISP[n^k, \log^m n]$. As such our hierarchy might be a useful tool in trying to separate these classes from $\NP$. Further as this hierarchy has a very syntactic definition, it lends itself to be used in areas such as bounded arithmetic. Bounded arithmetic are weak formal systems that are often used to model the reasoning needed to carry out complexity arguments. So even if $(\EXIST\SUNIV)^{\lin}_k$ basic properties don't immediately help one separate complexity classes they might be useful in bounded arithmetic to quantifier want kind of reasoning ability in terms of strengths of theories are needed to separate complexity classes.

Quantifiers such as $\SEXIST$ and $\SUNIV$ have been considered before in the context of linear time and quasi-linear in Bloch, Buss, Goldsmith~\cite{bbg98}. In particular, they look at a so-called sharply bounded hierarchy over quasi-linear time, using the notation $\tilde{\exists}$ and $\tilde{\forall}$ for sharply bounded quantifiers. Their class is contained in $\PTIME$ and has interesting closure properties, but it is unclear if it contains a well-known class like $\NLOGSPACE$. 

For this paper, we are interested in the structural and closure properties of $\EuLINH$ and the connections between $\EuLINH$ and the classes $\NLOGSPACE$, $\SC$, and $\NP$. As indicated above, one has  the containments $\NLOGSPACE, \SC \subseteq\EuLINH \subseteq \NP$. Since $\NLOGSPACE$ and $\SC$ are closed under complement one also has $\NLOGSPACE, \SC \subseteq \cup_k (\UNIV\SEXIST)^{\lin}_k \subseteq \coNP$. We first show some basic structural results such as if $(\EXIST\SUNIV)^{\lin}_k = (\EXIST\SUNIV)^{\lin}_{k+1}$ then $\NLOGSPACE \neq \NP$ and $\SC \neq \NP$. We further show if $(\EXIST\SUNIV)^{\lin}_1$ is contained in $(\UNIV\SEXIST)^{\lin}_k$ for some fixed $k$, then $\NP = \coNP$. We then consider  possible inclusions between our class and the classes $\NLOGSPACE$, $\SC$, and $\NP$. We show if 
$\NLOGSPACE$ or $\SC = \EuLINH$  then  $\NP = \coNP$, the polynomial time hierarchy is equal to the linear time hierarchy, and the linear time hierarchy is infinite. We then turn our attention to whether it is possible for $\EuLINH$ to equal $\NP$. We show that if $\EuLINH = \NP$, then the $\EuLINH$ hierarchy is infinite, $\LINH$ contains the boolean hierarchy over $\NP$, and either the linear time hierarchy is infinite or $\NP \neq \coNP$. Then we show that if the total sum of the length bounds on the universal quantifiers in $\EuLINH$ hierarchy is bounded to $k\log n$, then the restricted hierarchy, $\EuLINH_k$, is strictly contained in $\NP$. So in particular, if $\EuLINH_k$ contains $\NLOGSPACE$, which would happen if $\EuLINH_k=\EuLINH$, then $\NLOGSPACE \neq \NP$. It might be possible to further restrict the sizes of the universal quantifiers if the space requirements in the containment $(\EXIST\SUNIV)^{\lin}_{k}\subseteq  \NTISP[n^{k+1}, n] $ could be reduced. To this end we define a new quantifier $\xi$ and obtain partial results in this direction.

This paper is organized as follows: The next section contains the notations and main definitions used in this paper. This is followed by a section with some basic containments and closure properties of $\EuLINH$. Then \Nepom's Theorem is proven so as to connect it with $\EuLINH$.  This is followed by a short section reminding the reader of some useful padding and diagonalization techniques. The next two sections contain our main results, the first of these gives conditional results concerning what would happen if the $\EuLINH$ collapses, the second of these proves our results  concerning $\EuLINH$ and the classes $\NLOGSPACE$, $\SC$, and $\NP$.  The section which follows defines a new quantifier $\xi$ and a new hierarchy $\xiLINH$, and uses these to show a containment for $\EuLINH$ with a slightly lower space bound.  Finally, there is a conclusion with some suggestions for further investigation.

\section{Preliminaries}
A good survey of basic machine models and complexity classes can be found in Johnson~\cite{johnson90}. We briefly summarize here the models, classes and notation which will be used in this paper. We will use deterministic, nondeterministic, or alternating multi-tape Turing Machines as our basic machine models. These machines are assumed to have a read only input tape, a write only output tape, and some finite number of work tapes. Space is measured in terms of the total number of tape squares used on the work tapes. For the sake of simplicity, we assume without loss of generality that all languages are over the binary alphabet. We will often need at our disposal the ability to encode and decode pairs or finite sequences of strings.  To do this, the sequence of values $\langle x_1,\ldots , x_n\rangle$ --- when $n=2$, one has a pairing operation --- is defined to be the string obtained by replacing $0$Õs and $1$Õs in the $x_i$Õs by $00$ and $10$ respectively and by inserting a $01$ in between numbers. 

We write $\DTIME[t(n)]$ and $\NTIME[t(n)]$  for the languages decided by machines which are respectively deterministic or nondeterministic, and which run for at most $t(n)$ steps on inputs of length $n$. We write $\Sigma_k$-$\TIME[t(n)]$ and $\Pi_k$-$\TIME[t(n)]$ for the languages decided in $t(n)$ steps by machines which alternate at most $k-1$ times, in the former case the outermost alternation being existential, and in the latter case universal. We define $\DSPACE[s(n)]$ and $\NSPACE[s(n)]$ similarly except now the bound is on space rather than time. The classes $\DTISP[t(n),s(n)]$ and $\NTISP[t(n), s(n)]$ are the languages decided by machines  in $t(n)$ time steps using at most $s(n)$ work tape squares in respectively the deterministic or nondeterministic machine model.  For a language $L$, $\bar{L}$ is the complement of $L$, consisting of those strings $x$ over the alphabet that are not in $L$. Given a class of languages $\mathcal{C}$, we write $\co$-$\mathcal{C}$ for the languages $\{\bar{L} \, | \,L \in \mathcal{C}\}$.

Given these basic definitions, the following well-studied complexity classes can now be defined: $\DLIN := \DTIME[n]$, $\PTIME := \cup_k \DTIME[n^k]$,  $\NLIN := \NTIME[n]$, $\NP := \cup_k \NTIME[n^k]$, $\LOGSPACE := \DSPACE[\log n]$, $\NLOGSPACE := \NSPACE[\log n]$, and $\SC := \cup_k\cup_m \DTISP[n^k, \log^m n]$. It is known that
$\LOGSPACE \subseteq \NLOGSPACE = \coNLOGSPACE \subseteq \PTIME \subseteq \NP$
and that $\LOGSPACE \subseteq \SC \subseteq \PTIME$. As $\SC$, which stands for Steve's Class in honor of Steve Cook, is less-known we mention that $\SC$, in addition to the containments above, contains all the deterministic context free languages. Finally, we will be in interested in the linear and polynomial time hierarchies: $\LINH := \cup_j \Sigma_j$-$\TIME[n]$ and $\PH := \cup_j \SigmaP{j}$, where $\SigmaP{j} := \cup_k \Sigma_j$-$\TIME[n^k]$ and $\PiP{j} := \cup_k \Pi_j$-$\TIME[n^k]$. At the bottom levels of these hierarchies we have $\NLIN=\Sigma_1$-$\TIME[n]$, $\coNLIN=\Pi_1$-$\TIME[n]$ and $\SigmaP{1}=\NP$, $\PiP{1}=\coNP$ .
We also have trivially that $\Sigma_j$-$\TIME[n] \subseteq \Pi_{j+1}$-$\TIME[n]\subseteq \Sigma_{j+2}$-$\TIME[n]$ and
$\SigmaP{j} \subseteq \PiP{j+1} \subseteq \SigmaP{j+2}$, so  given our definition for each $j$,  $\Pi_{j}$-$\TIME[n]\subseteq \LINH$ and  $\PiP{j} \subseteq \PH$.

We consider strings $w$ built out of the quantifier alphabet  $\EXIST$, $\UNIV$, $\SEXIST$, and $\SUNIV$. Let $\tau$ be a collection of nondecreasing functions on the nonnegative integers. We will use $\tau$ to interpret quantifiers. The quantifier $\EXIST$ (resp\@., $\UNIV$) represents a quantifier of the form $\exists y\in\{0,1\}^{t(|x|)}$ (resp\@., $\forall y \in \{0,1\}^{t(|x|)}$) for some $t\in \tau$; the quantifier $\SEXIST$ (resp\@., $\SUNIV$) is supposed to
represent a quantifier of the form $\exists y \in\{0,1\}^{\log t(|x|)}$ (resp\@., $\forall y\in\{0,1\}^{\log t(|x|)}$). Given a class  $\mathcal{C}$ of languages, a string $w=w_k w_{k-1}\cdots w_0$ over our quantifier alphabet, and a set of functions $\tau$ as above, we define $(w\mathcal{C})^{\tau}$ by induction on $k$. If $k=0$ then $(w\mathcal{C})^{\tau} = \mathcal{C}$. For $k>0$, let $w' = w_{k-1}\cdots w_0$. Then one has four cases depending on the quantifier $w_k$: For instance, if $w_k$ is $\EXIST$, then $(w\mathcal{C})^{\tau}$ is the class of languages of the form $L = \{ x \,|\, \exists y \in \{0,1\}^{t(|x|)}$ $\langle x , y \rangle \in L' \}$ for some $t\in \tau$ and $L' \in (w'\mathcal{C})^{\tau} $. The other three cases are similarly defined using our way of interpreting $\UNIV$, $\SEXIST$, and $\SUNIV$ explained above. The main choices for $\tau$ we will be interested in are $\lin$ which consists of functions of the form $\ell(n) := k\cdot n$ for some $k>0$, and $\poly$ which consists of the polynomials $p(n) := n^k$ for some $k>0$. The principal starting choices for the complexity class $\mathcal{C}$ we will be interested in are $\DLIN$ and $\PTIME$. We will use two further ways to abbreviate our notation: (1) we will write $(w)^{\lin}$ for $(w\DLIN)^{\lin}$ and write $(w)^{\poly}$ for $(w\PTIME)^{\poly}$; (2) we will sometimes use subscripted notations such as $(w)_m$ as abbreviation for the string $\overbrace{w\cdots\cdots\cdots\cdots w}^{\mbox{$m$ concatentations}}$.  As an example of our notation scheme, the class $\NP = \SigmaP{1}$ could be written as $(\EXIST)^{\poly}$ and the class $\coNP=\PiP{1}$ could be written as $(\UNIV)^{\poly}$. The levels of the polynomial hierarchy for $k>1$ could be defined by saying $\SigmaP{k} =(\EXIST \PiP{k-1})^{\poly}$ and $\PiP{k} =(\UNIV \SigmaP{k-1})^{\poly}$. 
Finally, the new classes we will be interested in for this paper are $(\EXIST\SUNIV)^{\lin}_k := ((\EXIST\SUNIV)_k)^{\lin}$ and $(\UNIV\SEXIST)^{\lin}_k := ((\UNIV\SEXIST_k))^{\lin}$ for some integer $k \geq 0$. We will write $\EuLINH$ for the whole hierarchy $\cup_k (\EXIST\SUNIV)^{\lin}_k$ and $\UeLINH$ for the whole hierarchy $\cup_k (\UNIV\SEXIST)^{\lin}_k$.

\section{Basic Closure Properties and Containments}
\label{closure-section}
This section explores some of the basic relationships between, and closure properties of, our newly defined complexity classes $(\EXIST\SUNIV)^{\lin}_k$, $(\UNIV\SEXIST)^{\lin}_k$,  $\EuLINH$, $\UeLINH$. To begin we observe $(\EXIST\SUNIV)^{\lin}_k$ and $(\UNIV\SEXIST)^{\lin}_k$ are complements of each other. We also connect these classes to the levels of the linear time hierarchy.

\begin{prop}
\label{basic}
(a) For $k\geq 0$, $(\EXIST\SUNIV)^{\lin}_k = \co$-$(\UNIV\SEXIST)^{\lin}_k$. 
(b) For $k\geq 0$, $(\EXIST\SUNIV)^{\lin}_k \subseteq \Sigma_{2k}$-$\TIME[n]$ and
$(\UNIV\SEXIST)^{\lin}_k\subseteq \Pi_{2k}$-$\TIME[n]$. So both $\EuLINH$ and $\UeLINH$ are contained in $\LINH$.
\end{prop}
\begin{proof}
For (a), in the $k=0$ case, both classes are $\DLIN$. Further $\DLIN$ is closed under complement since given a machine $M$ for $L \in \DLIN$ we can swap its accept and reject states to obtain a machine for $\bar{L}$.  Now assume the proposition is true up to some $k \geq 0$, and let $t \in \lin$.  Remember that $x \in (\EXIST\SUNIV)^{\lin}_{k+1}$ iff $\exists y_1 \in \{0,1\}^{t(|x|)}$ such that $\forall y_2 \in \{0,1\}^{\log t(|x|)}$, $\langle x,y_1,y_2 \rangle \in (\EXIST\SUNIV)^{\lin}_k$.  Then it is clear that $x \in \co$-$(\EXIST\SUNIV)^{\lin}_{k+1}$ iff $\neg \exists y_1 \in \{0,1\}^{t(|x|)}$ such that $\forall y_2 \in \{0,1\}^{\log t(|x|)}$, $\langle x,y_1,y_2 \rangle \in (\EXIST\SUNIV)^{\lin}_k$.  By the induction hypothesis we have $\langle x,y_1,y_2 \rangle \not \in (\EXIST\SUNIV)^{\lin}_k$ iff $\langle x,y_1,y_2 \rangle \in (\UNIV\SEXIST)^{\lin}_k$; using that, and the fact that in first order logic $\neg \exists \forall$ is equivalent to $\forall \exists \neg$, we can say $x \in \co$-$(\EXIST\SUNIV)^{\lin}_{k+1}$ iff $\forall y_1 \in \{0,1\}^{t(|x|)}$, $\exists y_2 \in \{0,1\}^{\log t(|x|)}$ such that $\langle x,y_1,y_2 \rangle \in (\UNIV\SEXIST)^{\lin}_k$.  This shows that $x \in \co$-$(\EXIST\SUNIV)^{\lin}_{k+1}$ iff $x \in (\UNIV\SEXIST)^{\lin}_{k+1}$, so the induction step is proved.

To prove (b), we first note by increasing the size of the tape alphabet we can increase the number of computations that be done in $n$ steps by a constant factor (linear speed-up), so $\Sigma_{2k}$-$\TIME[O(n)] = \Sigma_{2k}$-$\TIME[n]$. Further for a language $L$ in  $(\EXIST\SUNIV)^{\lin}_k$ an alternating machine operating in time $O(n)$ can guess first the outermost quantifier, then universally guess the small universal quantifier, and so on,  with at most $2k$ alternations. Finally, it could simulate the innermost deterministic linear time machine used in deciding $L$. Thus, $(\EXIST\SUNIV)^{\lin}_k \subseteq \Sigma_{2k}$-$\TIME[n]$.
\end{proof}

\begin{prop}
\label{closure}
(a) $\LINH$ is closed under union, intersection, and complement.
(b) $(\EXIST\SUNIV)^{\lin}_k$, $(\UNIV\SEXIST)^{\lin}_k$, $\EuLINH$, and $\UeLINH$ are closed under union and intersection.
\end{prop}
\begin{proof}
$\LINH$ is closed under complement since if $L$ is in $\LINH$ then it is
in $\Sigma_k$-$\TIME[n]$ for some $k$, so $\bar{L}$ will be in $\Pi_k$-$\TIME[n]$. As we already observed in the preliminaries $\Pi_k$-$\TIME[n] \subseteq \Sigma_{k+1}$-$\TIME[n]\subseteq \LINH$. For both parts (a) and (b) of the proposition, the arguments for closure under union and intersection are essentially the same, so we indicate the proof for $(\EXIST\SUNIV)^{\lin}_k$. Let $L_1$ and $L_2$ be two $(\EXIST\SUNIV)^{\lin}_k$ languages. On input $x$ we can use pairing to guess pairs of strings for each $\EXIST$ or $\SUNIV$ quantifier, one of the two values to be used to check if $x$ is in $L_1$ and the other to check if $x$ is in $L_2$. This at most increases the size of the bound on each quantifier by a constant multiplicative factor, so this sequence of existential and universal guesses can still  be done in $(\EXIST\SUNIV)^{\lin}_k$. The deterministic machine for the union, first unpacks each pair to produce a string to start simulating the $\DLIN$ machine for $L_1$. If this simulation accepts then the union machine accepts; otherwise, the other components of each pair are extracted and the machine simulates the $\DLIN$ machine for $L_2$. If it accepts, then the union machine accepts; otherwise it reject. The total runtime will be at most the sum of the two runtimes so still linear. For intersections the same sort of algorithm is done except that the machine always performs the simulation of the $\DLIN$ machine for $L_1$ followed by the $\DLIN$ machine for $L_2$ and only accepts if both accepted.
\end{proof}

\begin{prop}
\label{basic2}
(a) $\NLIN \subseteq (\EXIST\SUNIV)^{\lin}_1$ and $\coNLIN \subseteq (\UNIV\SEXIST)^{\lin}_1$.
(b) For $k\geq 0$, $(\EXIST\SUNIV)^{\lin}_k\subseteq \NTISP[n^{k+1}, n] \subseteq \NTIME[n^{k+1}]$, so $\EuLINH$ is contained in $\NP$ and $\UeLINH$ is contained in $\coNP$.
(c) For $k\geq 1$, $(\EXIST\SUNIV)^{\poly}_k = \NP$. 
\end{prop}
\begin{proof}
For (a) notice that using an $\EXIST$ quantifier to guess a string $w$ of nondeterministic moves of an $\NLIN$ machine $M$ on input $x$, a deterministic linear time machine could then use these choices together with $x$ to decide $M$'s language. This show $\NLIN \subseteq (\EXIST)^{\lin} \subseteq (\EXIST\SUNIV)^{\lin}_1$. By taking complements we have the $\coNLIN$ result.

To prove (b) we show by induction on $k$ that  $(\EXIST\SUNIV)^{\lin}_k$ is contained in $ \NTISP[n^{k+1},n]$. As $\EuLINH = \cup_k (\EXIST\SUNIV)^{\lin}_k$ and $\NP = \cup_k  \NTIME[n^{k+1}]$, this implies $\EuLINH$ is contained in $\NP$. The fact that $\UeLINH \subseteq \coNP$ then follows from Proposition~\ref{basic}~$(a)$. For the $k=0$ case, we obviously have $\DLIN \subseteq \NLIN = \NTISP[n,n]$. So assume the statement is true up to some $k \geq 0$.  Let $L$ be a language in $(\SUNIV(\EXIST\SUNIV)_k)^{\lin}$. We will argue $L$ is contained in $ \NTISP[n^{k+2}, n]$. By the definition of $L$ there must be some $(\EXIST\SUNIV)^{\lin}_k$ languages $L'$ such that $x \in L$ if and only if $\forall y \in \{0, 1\}^{\log t(|x|)}, \langle x,y\rangle \in L'$. Here $t(n)$ is some growth rate from $\lin$ so is of the form $m\cdot|x|$.  By our induction hypothesis there is a $\NTISP[n^{k+1}, n]$ machine $M'$ which decides $L'$. Let $M$ be the machine which cycles through the $m\cdot|x|$ strings $y$ of length $\log (m\cdot|x|)$, reusing the space. For each $y$, $M$ then simulates $M'$ on $\langle x, y\rangle$. If $M'$ ever rejects, then $M$ rejects. Otherwise, if $M'$ accepts all such strings, then $M$ accepts. This involves less than a linear amount of additional space. Further, $M$'s runtime is $O(m\cdot|x|)$ times longer than the runtime of $M'$, so using linear speed-up, $M$ will operate in time $n^{k+2}$. Further, if $x \in L$ then for every string $y$ with $|y| \leq \log (m\cdot|x|)$,
$\langle x, y \rangle \in L'$, so there must be some path making $M'$ accept. These accepting paths of $M'$ for each $y$, taken together, can be used to make an accepting path for $M$. On the other hand, if $x\not\in L$, then for some $y$,  there is no path which makes $M'$ accepts $\langle x, y\rangle$, and so when this $y$ is checked by $M$, it too will reject. Thus, we have shown  $\SUNIV(\EXIST\SUNIV)^{\lin}_k$ is contained in $\NTISP[n^{k+2},n]$. But then $(\EXIST\SUNIV(\EXIST\SUNIV)_k)^{\lin} = (\EXIST\SUNIV)^{\lin}_{k+1}$ will also be in $\NTISP[n^{k+2},n]$ since given a language in $(\EXIST\SUNIV)^{\lin}_{k+1}$, an $\NTISP[n^{k+2},n]$ machine could in nondeterministic linear time and space guess the outermost existential quantifier and then run the $\NTISP[n^{k+2},n]$ machine for the corresponding language in $(\SUNIV(\EXIST\SUNIV)_k)^{\lin} $. Hence, the induction step is proved and the result follows.

For (c), we first note by the same argument as in (a), we have $\NP \subseteq (\EXIST)^{\poly}\subseteq (\EXIST\SUNIV)^{\poly}_k$ for $k\geq 1$. To see $(\EXIST\SUNIV)^{\poly}_k \subseteq \NP$, we can use the same algorithm as in (b). Guessing an $\EXIST$ quantifier will now involve guessing a string of polynomial length, but this can be handle in $\NP$; simulating a $\SUNIV$ quantifier will involve cycling over the at $|x|^m$ string of length $\log t(|x|)$ where $t$ is a polynomial of the form $|x|^m$ for some $m$, so can be done in polynomial time. 
\end{proof}

Given the result (b) above it is reasonable to ask if $(\EXIST\SUNIV)^{\lin}_k = \NTISP[n^{k+1}, n]$ or if $(\EXIST\SUNIV)^{\lin}_k =\NTIME[n^{k+1}]$ for $k>0$? We conjecture the latter is not the case and we will give some evidence for this in Corollary~\ref{infinite3}. Nevertheless, the results of the next section can be viewed as telling us that $(\EXIST\SUNIV)^{\lin}_k$ might be close to $\NTISP[n^{k+1}, n]$.

\section{\Nepom's Theorem}
\begin{thm}
\label{nepom}
For $k\geq 1$ and $0\leq\epsilon < 1$,
$\NTISP(n^{k(1-\epsilon)}, n^\epsilon)$ is contained in $(\EXIST\SUNIV)^{\lin}_{k}$.
\end{thm}
\begin{proof}
Our proof of \Nepom's Theorem is essentially the same as found in say {\Hajek } and \Pudlak~\cite{hp} or Fortnow et al\@.~\cite{flmv06}. Here, though, we emphasize the details that imply our containment. 
Consider a  $k$-tape Turing Machine $M$ for a language $L$ in $\NSPACE( n^\epsilon)$.  To make our argument easier, we will assume that when $M$ halts, it erases its tapes, moves to the left-most square on each tape, and  then either halts accepting or rejecting.  A configuration $C$ of $M$ is $k+3$ tuple $\langle q, i, c,  t_1, \ldots, t_k \rangle$ representing the state of $M$, the index of the square being read on the input tape, the character being written to the output tape,  and the strings representing the visited contents of each work tape. The position of work tape heads are indicated in each $t_i$ by using an alphabet symbol with an underscore beneath it for that tape square.

Consider the language $L_k$ consisting of 4-tuples of the form $(x, C_s, C_t, 1^{|x|^{k(1-\epsilon)}})$ such that  $C_s$ and $C_t$ represent configurations of $M$ on input $x$, and there is some $|x|^{k(1-\epsilon)}$-step computation of $M$ beginning in configuration $C_s$ and ending in configuration $C_t$. We proceed by induction on $k$ and argue that $L_k$ is in $(\EXIST\SUNIV)^{\lin}_{k}$. When $k=1$, we need to argue that $L_1$ is contained in  $(\EXIST\SUNIV)^{\lin}_{1}$. Verifying that $C_s$ and $C_t$ are indeed configurations of $M$ can be done in linear time in $C_s$ and $C_t$.  Using the $\EXIST$ quantifier we can guess a sequence $C_1, \ldots, C_v$ of configurations of $M$  where $C_s = C_1$, $C_t=C_v$, $v=|x|^{1-\epsilon}$. This is possible since each configuration has size $|x|^{\epsilon}$ and   $v=|x|^{1-\epsilon}$, so the string we are guessing has length $O(|x|)$. Using a $\SUNIV$ quantifier to guess every $i \in \{1, \ldots, v-1\}$, one can then verify in deterministic linear time that (for each $i$) $C_{i+1}$ follows from $C_i$ according to a transition in $M$, or, if $C_i$ is a halted configuration, that $C_i = C_{i+1}$. This shows the result holds for $k=1$.

Assume $L_k \in (\EXIST\SUNIV)^{\lin}_{k}$ holds for $k\geq 1$ and consider an instance  $(x, C_s, C_t, 1^{|x|^{(k+1)(1-\epsilon)}})$ of  $L_{k+1}$. Again, with an $\EXIST$ quantifier we can guess a sequence $C_1, \ldots, C_v$ of configurations of $M$  where $C_s = C_1$, $C_t=C_v$, $v=|x|^{1-\epsilon}$.  Then using a $\SUNIV$ quantifier we can check if for each $i = 1, \ldots v-1$, that we can guess the 4-tuple $(x, C_i, C_{i+1}, 1^{|x|^{k(1-\epsilon)}})$ and verify that it is in $L_k$. This shows $L_{k+1}$ is in $(\EXIST\SUNIV\EXIST(\EXIST\SUNIV)_k)^{\lin}$, but $(\EXIST(\EXIST\SUNIV)_k)^{\lin}=(\EXIST\SUNIV)^{\lin}_{k}$ because we can always combine outer two existential quantifiers of linear size into one existential quantifier still of linear size.  Hence, we have $L_{k+1}$ is in $(\EXIST\SUNIV)^{\lin}_{k+1}$ and the induction step is proved.

Now if $M$ was for a language $L$ that was not only in  $\NSPACE[n^\epsilon]$ but in $\NTISP[n^{k(1-\epsilon)}, n^\epsilon]$, then on input $x$ we could guess the 4-tuple $(x, C_s, C_h, |x|^{k(1-\epsilon)})$, where $C_s$ is the starting configuration of $M$ on $x$ and $C_h$ is the unique possible accepting, halting configuration; and check if this 4-tuple is in $L_k$. This would show $L \in \EXIST(\EXIST\SUNIV)_k)^{\lin}$, but as we have just argued, $(\EXIST(\EXIST\SUNIV)_k)^{\lin}=(\EXIST\SUNIV)^{\lin}_{k}$. Hence, the theorem follows.
\end{proof}

Combining Proposition~\ref{basic2} and Theorem~\ref{nepom} above, we have an almost optimal containment
for the levels of our hierarchy:
\begin{cor}
For $k\geq 1$ and $0\leq\epsilon < 1$,
$\NTISP(n^{k(1-\epsilon)}, n^\epsilon)\subseteq(\EXIST\SUNIV)^{\lin}_{k}\subseteq  \NTISP[n^{k+1}, n] $.
\end{cor}
The next corollary also follows from Theorem~\ref{nepom}  since $\NLOGSPACE = \cup_k \NTISP(n^k, \log n)$ and $\SC  := \cup_k\cup_m \DTISP[n^k, \log^m n]$.
 \begin{cor}
(a) $\NLOGSPACE$ is contained in $\EuLINH$.
(b) $\SC$ is contained in  $\EuLINH$.
\end{cor}

\section{Common Arguments}
In this section, we briefly present well-known results concerning padding and time hierarchies which we will need for our later results. To begin we say a function $t(n)$ is {\em time constructible} if there is a $t(n)$ time bounded Turing Machine $M$ such that for each $n$, $M$ runs for exactly $t(n)$ steps on an input of length $n$. Given a complexity class $\mathcal{C}$, we write $R_{\pad}(\mathcal{C})$ for the class of languages $L$ for which there is a logspace computable, time constructible function $t(n) \geq n$, such that $x\in L$ if and only if $x0^i \in L'$ where $L' \in \mathcal{C}$ and $|x0^i| = t(n)$. The string $x0^i$ here is the string $x$ padded by a string of $0$'s of length $i$. The next lemma collects together the main results we need about the padded versions of the complexity classes of this paper.

\begin{lem}
\label{padding}
(a) Let $\mathcal{C}_1 \subseteq \mathcal{C}_2$ be classes of languages. Then $R_{\pad}(\mathcal{C}_1) \subseteq R_{\pad}(\mathcal{C}_2)$.
(b) $R_{\pad}(\NLIN)=R_{\pad}(\EuLINH)=R_{\pad}(\NP) = \NP$.
(c) $R_{\pad}(\co$-$\NLIN)=R_{\pad}(\UeLINH)=R_{\pad}(\coNP) = \coNP$.
\end{lem}
\begin{proof}
(a) If $L$ is a language in $R_{\pad}(\mathcal{C}_1)$. Then there must be some language $L'$ in $\mathcal{C}_1$ and a time constructible $t(n)$ such that $x \in L$ if and only if $x0^i \in L'$ where $|x0^i|=t(n)$.  Since $\mathcal{C}_1 \subseteq \mathcal{C}_2$, this same $L'$ and $t$ show $L$ is in  $R_{\pad}(\mathcal{C}_2)$.

(b) As $\NLIN \subseteq \EuLINH \subseteq \NP$ by (a) it suffices to show $R_{\pad}(\NP) \subseteq \NP$ and $\NP \subseteq R_{\pad}(\NLIN)$. The first inclusion follows since given a language $L \in  R_{\pad}(\NP)$ and an input $x$ a nondeterministic Turing machine can compute the logspace computable, time constructible function $t(n)$ for  $L$ to get $x0^i$. It can then simulate the $\NP$ machine for this language on $x0^i$. The total runtime will be bounded by a composition of polynomials, showing $L$ is in $\NP$. For the second inclusion, let $L\in \NP$ be decidable in time bounded by a polynomial $p(n)$. Then the language $\{x0^{p(|x|)-|x|} \, | \, x\in L\}$ will be in $\NLIN$ and show $L \in R_{\pad}(\NLIN)$.

The argument for (c) is essentially the same as for (b), except that the inclusions shown are $R_{\pad}(\coNP) \subseteq \coNP$ and $\coNP \subseteq R_{\pad}(\co$-$\NLIN)$.
\end{proof}

We next turn to two diagonalization results for time complexity classes which we will use but not prove. The first of these is due to Chandra and Stockmeyer~\cite{cs76}, which Williams~\cite{williams2005} gives the catchy name: The No Complementary Speedup Theorem. The second of these is the Nondeterministic Time Hierarchy Theorem~\cite{sfm78, zak83}.
 
\begin{thm}
\label{diag1}
Let  $k$ be a positive integer and $t$ a time constructible function. Then $\Pi_k$-$\TIME[t] \not\subseteq \Sigma_k$-$\TIME[o(t)]$. 
\end{thm}

\begin{thm}
\label{diag2}
Let $t_1(n)$ and $t_2 (n)$ be functions with $t_2(n)$ time constructible. If  $t_1(n + 1) \in o(t_2(n))$ then 
$\NTIME[t_1 ] \subsetneq \NTIME[t_2 ]$.
\end{thm}
\section{Levels of $\EuLINH$}
In this section we present consequences of $\EuLINH$ being contained in $\UeLINH$ and we consider the question of whether or not the $\EuLINH$ hierarchy is infinite. 
\begin{thm}
If $(\EXIST\SUNIV)^{\lin}_1$ is contained in $(\UNIV\SEXIST)^{\lin}_k$ then $\NP=\coNP$.
\end{thm}
\begin{proof}
As $\NLIN \subseteq (\EXIST\SUNIV)^{\lin}_1$, this implies $\NLIN \subseteq (\UNIV\SEXIST)^{\lin}_k$, and thus $\NLIN \subseteq \UeLINH$. So by Lemma~\ref{padding} we have $\NP = R_{\pad}(\NLIN) \subseteq R_{\pad}(\UeLINH) = \coNP$, and hence $\NP = \coNP$
\end{proof}

\begin{thm}
Let $k>0$. If $(\EXIST\SUNIV)^{\lin}_k = (\EXIST\SUNIV)^{\lin}_{k+1}$ then $\NLOGSPACE \neq \NP$ and
$\SC \neq \NP$.
\end{thm}
\begin{proof}
First, if  $(\EXIST\SUNIV)^{\lin}_k = (\EXIST\SUNIV)^{\lin}_{k+1}$, then 
$$(\EXIST\SUNIV)^{\lin}_{k+2}= (\EXIST\SUNIV(\EXIST\SUNIV)_{k+1})^{\lin} = (\EXIST\SUNIV(\EXIST\SUNIV)_k)^{\lin} = (\EXIST\SUNIV)^{\lin}_k.$$
Continuing up the hierarchy in this fashion one thus has $(\EXIST\SUNIV)^{\lin}_k = \EuLINH$. By Proposition~\ref{basic2} (b), this implies $\EuLINH \subseteq \NTIME[n^{k+1}] \subsetneq \NP$. The result then follows, as by Theorem~\ref{nepom}, we know $\SC\subseteq \EuLINH$ and $\NLOGSPACE \subseteq \EuLINH$.
\end{proof}

\section{Other complexity classes and $\EuLINH$}
In this section we investigate various consequences of certain inclusions holding among $\EuLINH$, $\LINH$, $\SC$, $\NLOGSPACE$, and $\NP$.
\begin{prop}
\label{infinite}
If $\LINH = \PH$ then for all $k$, $\LINH \neq \Sigma_k$-$\TIME[n]$.That is, the linear time hierarchy is infinite.
\end{prop}
\begin{proof}
By Theorem~\ref{diag1} we know that   $\Pi_k$-$\TIME[n^2] \not\subseteq \Sigma_k$-$\TIME[n]$. We also know $\Pi_k$-$\TIME[n^2]\subseteq \PiP{k} \subseteq \SigmaP{k+1}$. Therefore, if $\LINH =  \Sigma_k$-$\TIME[n]$ for some $k$, then it is strictly contained in $\PH$.
\end{proof}

\begin{cor}
If $\coNP \subseteq \EuLINH$ then for all $k$, $\LINH \neq \Sigma_k$-$\TIME[n]$. 
\end{cor}
\begin{proof}
Proposition~\ref{basic2} (b) and $\coNP \subseteq \EuLINH$ would imply $\coNP \subseteq  \EuLINH \subseteq \NP$. Hence, $\NP=\coNP=\EuLINH$.
Further as $\NP = \coNP$ implies $\NP = \PH$ and $\EuLINH \subseteq \LINH \subseteq \PH$, we would have $\LINH=\PH$ and so the result follows from Proposition~\ref{infinite}.
\end{proof}

For the next result recall that the boolean hierarchy above $\NP$ is defined as $\BH := \cup_i \BH_i$, where
$\BH_1 = \NP$, where $\BH_{2i}$ is the class of languages consisting of the intersection of  a language in $\BH_{2i-1}$ with a language in $\coNP$, and where $\BH_{2i+1}$ is the class of languages consisting of the intersection of  a language in $\BH_{2i}$ with a language in $\NP$. Kadin~\cite{kadin88} has shown if this hierarchy collapses so does the polynomial hierarchy. We next consider some consequences of $\EuLINH$ being equal to $\NP$.
\begin{thm}
\label{infinite2}
If $\EuLINH = \NP$ then: 
(a) For all $k\geq 0$, $\EuLINH \neq (\EXIST\SUNIV)^{\lin}_k$.
(b) $\LINH$ contains $\BH$.
(c) Either for all $k$, $\LINH \neq \Sigma_k$-$\TIME[n]$ or $\NP \neq \coNP$.
\end{thm}
\begin{proof}
To see (a) notice Proposition~\ref{basic2} (b) implies $(\EXIST\SUNIV)^{\lin}_k \subseteq \NTIME[n^{k+1}] \subsetneq \NP$, where $\NTIME[n^{k+1}] \subsetneq \NP$ follows by the nondeterministic time hierarchy theorem (Theorem~\ref{diag2}).

For (b) observe that if $\EuLINH = \NP$ then by Proposition~\ref{basic2} (b)  we have  $\NP=\EuLINH\subseteq \LINH$. The result then follows as $\LINH$ is closed under unions, intersections, and complements.

Lastly, for (c) suppose $\EuLINH = \NP$, $\NP = \coNP$,  and  $\LINH = \Sigma_k$-$\TIME[n]$ for  some $k$.  Then $\EuLINH = \NP =\coNP = \PH$, as $\EuLINH \subseteq \LINH \subseteq \PH$ we have $\LINH=\PH$. Thus, $\LINH = \Sigma_k$-$\TIME[n]$ contradicts Proposition~\ref{infinite}.
\end{proof}

At the end of Section~\ref{closure-section}, we said it was unlikely that $(\EXIST\SUNIV)^{\lin}_m = \NTIME[n^{m+1}]$. The next corollary gives some consequences of this equality.
\begin{cor}
\label{infinite3}
If for all $m'>0$, there is a $m>m'$ such that $(\EXIST\SUNIV)^{\lin}_m = \NTIME[n^{m+1}]$, then:
(a) For all $k\geq 0$, $\EuLINH \neq (\EXIST\SUNIV)^{\lin}_k$.
(b) $\LINH$ contains $\BH$.
(c) Either for all $k$, $\LINH \neq \Sigma_k$-$\TIME[n]$ or $\NP \neq \coNP$.
\end{cor}
\begin{proof}
The hypothesis above implies $\EuLINH = \NP$ as there are unboundedly large $m$ satisfying   $(\EXIST\SUNIV)^{\lin}_m = \NTIME[n^{m+1}]$, so  the union $\cup_m (\EXIST\SUNIV)^{\lin}_m$ for these $m$'s is  $\EuLINH$ and the union $\cup_m  \NTIME[n^{m+1}]$ is $\NP$. The result then follows from Theorem~\ref{infinite2}.
\end{proof}

\begin{thm}
\label{collapses}
(a) If $\EuLINH \subseteq \coNP$ then $\NP= \PH$.
(b) If $\EuLINH = \LINH$ then $\NP= \PH$.
(c) If $\SC =\EuLINH$ or $\NLOGSPACE = \EuLINH$ 
then $\EuLINH=\LINH=\NP= \PH$.
\end{thm}
\begin{proof}
For (a), if $\EuLINH \subseteq \coNP$, then in particular $\NLIN \subseteq \coNP$ and so by padding
$\NP \subseteq \coNP$. Thus, $\NP =\coNP = \PH$.

To see part (b) notice that $\LINH$ contains $\co$-$\NLIN$, so if $\EuLINH = \LINH$ then $\co$-$\NLIN \subseteq \EuLINH$.
Thus, by Lemma~\ref{padding}, we have $\coNP \subseteq \NP$. Hence $\NP = \coNP = \PH$.

For (c) let $\mathcal{C}$ be either $\SC$ or $\NLOGSPACE$. In either case, $\mathcal{C}$ is closed under complement and closed under log space (and hence, polynomial length padding) reductions~\cite{immerman88, sz87}.  So if $\mathcal{C} = \EuLINH$ then in particular
$\NLIN \subseteq \mathcal{C}$ and by Lemma~\ref{padding} we get $\NP \subseteq \mathcal{C}$. Since in both cases, we know $ \mathcal{C}  \subseteq \NP$, we get $\mathcal{C}=\NP$. So by closure under complement of $\mathcal{C} = \NP = \coNP = \PH$. As  we are assuming $\mathcal{C} = \EuLINH\subseteq \LINH \subseteq \PH$, we also have $\mathcal{C} = \EuLINH = \LINH = \PH$. 
\end{proof}

The argument for part (c) thus shows:
\begin{cor}
\label{collapses2}
If $\SC = \NP$ or $\NLOGSPACE = \NP$ then $\EuLINH=\LINH = \NP= \PH$.
\end{cor}

If the conclusion were to happen then by Proposition~\ref{infinite} and  Theorem~\ref{infinite2}, both $\LINH$ and $\EuLINH$ must not collapse.  

\begin{cor}
For $k\geq 1$, if $\EuLINH \subseteq \Sigma_k$-$\TIME[n]$ then $\NLOGSPACE \neq \NP$ and $\SC \neq \NP$.
\end{cor}
\begin{proof}
By the previous corollary, $\SC = \NP$ or $\NLOGSPACE = \NP$ implies $\EuLINH=\LINH = \PH$. By Proposition~\ref{infinite} we then have $\LINH \neq \Sigma_k$-$\TIME[n]$ $\forall k \geq 1$. Since $\Sigma_k$-$\TIME[n] \subseteq \LINH$ by definition, we have $\LINH \not \subseteq \Sigma_k$-$\TIME[n]$, and thus $\EuLINH \not \subseteq \Sigma_k$-$\TIME[n]$.
\end{proof}\\

Intuitively, $\EuLINH=\LINH$ comes very close to showing that $\LINH$ collapses. We already know by Theorem~\ref{collapses} (b) that in itself this condition implies $\NP=\PH$. So if $\EuLINH=\LINH$ does imply $\LINH$ collapses, then together with the consequence  $\NP=\PH$, this would force us to conclude the premise of Corollary~\ref{collapses2} is false.  i.e., $\SC \neq \NP$ and $\NLOGSPACE \neq \NP$. In an attempt to give at least some justification for this intuition that $\EuLINH$ is a small subclass of $\LINH$, one might consider a slightly weaker version of $\EuLINH$ with smaller sharply bounded universal quantifiers and see if it can equal $\NP$ or not. Let $s\in O(\log n)$. Extending our notation slightly we will write $\SUNIV^s$ in a quantifier string to mean the same as $\SUNIV$ except where the bound on the quantifier is $s(n)$ rather $k\log n$.  Consider the class
$\mathcal{C}_{m,k} := (\EXIST \SUNIV^{s_m} \cdots \EXIST \SUNIV^{s_1})^{\lin}$ where $(\sum_{j=1}^m s_j(n)) \leq k \log n$. Let $\EuLINH_k := \cup_m \mathcal{C}_{m,k}$. So $\EuLINH_k$ is analogous to $\EuLINH$, except that we have forced the sharply bounded quantifiers to be slightly smaller. As an example, if each $s_1, \ldots s_m$ were $\log\log n$, then for each $m\geq 1$, $(\EXIST \SUNIV^{s_m} \cdots \EXIST \SUNIV^{s_1})^{\lin}$ would be contained in $\EuLINH_1$.  The next result shows if we could tighten \Nepom's result to show $\SC$ and $\NLOGSPACE$ are in $\EuLINH_k$ for any $k \geq 1$ then $\SC \neq \NP$ and $\NLOGSPACE \neq \NP$.

\begin{thm}
For $k\geq 1$, $\EuLINH_k\subseteq\NTIME[n^{k+1}] \subsetneq \NP$.
\end{thm}
\begin{proof}
Suppose $L$ is in $\EuLINH_k$ and hence in $\mathcal{C}_{m,k}$ for some $m$. Let $s_1(n), \ldots s_m(n)$ be the size bounds on the sharply bounded quantifiers. By the definition of $\EuLINH_k$, we have $(\sum_{j=1}^m s_j(n)) \leq k \log n$.  If one examines the proof of Proposition~\ref{basic}, the run time of the $\NP$ machine given there to simulate $L$ would be
$$O(n +2^{s_m}(n +2^{s_{m-1}}( \cdots (n+2^{s_1}\cdot n)))),$$ 
the $n$'s coming from guessing the existential quantifiers, the $2^{s_i}$'s coming from looping over all the choices of strings of length $s_i$. This run time can be rewritten as
$$O(n(1+2^{s_m} + 2^{s_m+s_{m-1}} + \cdots + 2^{(\sum_{j=1}^m s_j(n))} )).$$
which can be bounded by $O(n\cdot m 2^{k \log n})= O(n^{k+1})$, the latter equality following as $m$ is constant. Thus, the result follows.
\end{proof}

\section{Reducing the Space Requirements for $\EuLINH$}
With the goal of reducing the space requirements needed for our upper bound on $\EuLINH$, we define a new quantifier, and thus a new hierarchy.  Given a class $\mathcal{C}$, define $\xi \mathcal{C}$ to be the class of languages of the form $\lbrace x \, \vert \, \exists w \in \lbrace 0,1 \rbrace^{\vert x \vert^{d_1}} \, \forall i < \vert x \vert^{d_2}, \langle x, i, (w)_{f_1(i)}, (w)_{f_2(i)} \rangle \in L \rbrace$ for some $L \in \mathcal{C}$, some constants $d_1$ and $d_2$ (where $d_1 \geq d_2$), and some $f_1, f_2: [0,\vert x \vert^{d_2}-1] \rightarrow [0,\vert x \vert^{d_2}-1]$.  Note that $(w)_i$ denotes the $i$th block of $w$ when $w$ is split into $\vert x \vert^{d_2}$ blocks, each of size $\vert x \vert^{d_1 - d_2}$, and the functions $f_1$ and $f_2$ are used to select blocks of $w$.  With this new quantifier $\xi$, we can define $(\xi)^{\lin}_k$ and $\xiLINH$ in exactly the same way that we defined $(\EXIST \SUNIV)^{\lin}_k$ and $\EuLINH$.  Notice that, like in $(\EXIST \SUNIV)^{\lin}_k$, the universal quantifiers of $(\xi)^{\lin}_k$ are of logarithmic size, though the existential quantifiers are of polynomial rather than linear size.\\
\\
We first show the relationship between the two hierarchies.

\begin{thm}
\label{xi-versus-Eu}
For $k \geq 0$, $(\EXIST \SUNIV)^{\lin}_k \subseteq (\xi)^{\lin}_k$.
\end{thm}
\begin{proof}
The proof is by induction.  When $k=0$, we have $(\EXIST \SUNIV)^{\lin}_k = \DLIN = (\xi)^{\lin}_k$, so the theorem holds.  Now assume the theorem holds up to some $k-1$.  Let $L$ be a language in $(\EXIST \SUNIV)^{\lin}_k$.  Then there is an $L' \in (\EXIST \SUNIV)^{\lin}_{k-1} \subseteq (\xi)^{\lin}_{k-1}$ such that $L = \lbrace x \, | \, \exists y_1 \in \lbrace 0,1 \rbrace^{c_1 |x|} \, \forall y_2 \in \lbrace 0,1 \rbrace^{\log(c_2 |x|)}, \langle x, y_1, y_2 \rangle \in L' \rbrace$ for some constants $c_1,c_2$. We shall define a $(\xi)^{\lin}_k$-machine $M$ that decides $L$.

On input $x$, for $M$'s outermost $\xi$ quantifier, the guessed $w$ has length $c_1 c_2 |x|^2$ (which is within the polynomial bound), $i$ ranges from 0 to $c_2 |x| - 1$ (also within the polynomial bound), and we have $f_1(i)=i$, $f_2(i)=$ min$(i+1, c_2 |x| - 1)$.  ($f_1$ and $f_2$ were chose in this way to pick consecutive blocks from $w$, but to pick the same block when $i$ is at its maximum value.)  Note that the set of all $i$s, when written in binary, is exactly $\lbrace 0,1 \rbrace^{\log(c_2 |x|)}$.  Each block of $w$ is of length $c_1 |x|$, and in fact $w$ is meant to represent $c_2 |x|$ concatenations of the string guessed for $y_1$ in the definition of $L$.

Given the above definitions, $M$ operates on strings of the form $\langle x, i, (w)_i, (w)_{i+1} \rangle$, where we understand $(w)_{i+1}$ to mean $(w)_i$ when $i$ is at its maximum value.  $M$'s first step is to verify that $(w)_i = (w)_{i+1}$; this is done to ensure that $w$ is indeed $c_2 |x|$ concatenations of the same string.  This check can be done in linear time.  Assuming that that check passes, $M$ then ``renames'' its inputs by saying $y_1 = (w)_i$ and $y_2 = i$ (in binary), and checks if $\langle x, y_1, y_2 \rangle \in L'$ by using the $(\xi)^{\lin}_{k-1}$-machine for $L'$.

To verify $M$'s correctness, we need only consider the branches where the guessed $w$ is $c_2 |x|$ concatenations of the same string, since we know $M$ will reject on all existential branches where this is not the case.  If $x \in L$, then there is some fixed $y_1$ such that $\langle x, y_1, y_2 \rangle \in L'$ for each $y_2 \in \lbrace 0,1 \rbrace^{\log(c_2 |x|)}$, and so $M$ will accept $x$ when the outer $\xi$ chooses the $w$ which consists of repeated $y_1$s.  Conversely, if $x \not \in L$, then for all $w$ which consist of concatenations of the same $y_1$, there is some $y_2 \in \lbrace 0,1 \rbrace^{\log(c_2 |x|)}$ such that $\langle x, y_1, y_2 \rangle \not \in L'$, and so $M$ will reject $x$.
\end{proof}

\begin{cor}
$\EuLINH \subseteq \xiLINH$.
\end{cor}

\noindent Theorem~\ref{xi-versus-Eu} also gives us an analogous result to \Nepom's Theorem for $\xiLINH$.

\begin{cor}
For $k\geq 1$ and $0\leq\epsilon < 1$, $\NTISP(n^{k(1-\epsilon)}, n^\epsilon)$ is contained in $(\xi)^{\lin}_{k}$.
\end{cor}

\noindent In order to prove an upper bound on this new hierarchy, we now show that the $\xi$ quantifiers are collapsable.

\begin{thm}
\label{xi-collapse}
For any class $\mathcal{C}$, $\xi \xi \mathcal{C} \subseteq \xi \mathcal{C}$.
\end{thm}
\begin{proof}
First, let us be precise about the definition of $\xi \xi \mathcal{C}$.  A machine $M$ which decides a language $L \in \xi \xi \mathcal{C}$ computes the following on input $x$: $$\exists w \in \lbrace 0,1 \rbrace^{\vert x \vert^{d_1}} \, \forall i < \vert x \vert^{d_2} \, \exists w' \in \lbrace 0,1 \rbrace^{\vert x \vert^{d'_1}} \, \forall i' < \vert x \vert^{d'_2},$$ $$R(x, i, (w)_{f_1(i)}, (w)_{f_2(i)}, i', (w')_{f'_1(i')}, (w')_{f'_2(i')})$$ where $R$ is a predicate computable within the bounds of the class $\mathcal{C}$.  In order to compress the two $\xi$s into one, we define the string $v$ to be $\langle w'_0, w'_1, \ldots, w'_{\vert x \vert^{d_2}-1} \rangle$, where $w'_n$ is the string guessed by the inner $\xi$ when the $i$ (from the outer $\xi$) is equal to $n$.  Then, we can rewrite the quantifier string $\exists w \forall i \exists w' \forall i'$ as $\exists w \exists v \forall i \forall i'$.  Now, we use the following notation for our new ``compressed'' $\xi$: $$\exists y \in \lbrace 0,1 \rbrace^{\vert x \vert^{e_1}} \, \forall j < \vert x \vert^{e_2}, S(x, j, (y)_{g_1(j)}, (y)_{g_2(j)})$$ To show that this is the same, we will define the new existential string $y$ to contain all of $w$ and $v$, though $y$ will not be a simple concatenation of these two strings.  As we will see shortly, the size of $y$ is $2(|x|^{d_2 + d'_2})(|x|^{d_1-d_2} + |x|^{d'_1 - d'_2})$, and so we choose $e_1$ so that the bound on $y$, $\vert x \vert^{e_1}$, is bigger than this quantity.  Each block of $y$ contains one block of $w$ and one block of \textit{one} of the $w'_k$s, so each block has size $\vert x \vert^{d_1-d_2} + \vert x \vert^{d'_1-d'_2}$.  And thus, we choose $e_2$ such that $|x|^{e_2} \geq 2(|x|^{d_2 + d'_2})$.

For the string $y$, we define it as follows.  The first $2 \vert x \vert^{d'_2}$ blocks contain the necessary information when $i = 0$ and $i'$ ranges from 0 to $\vert x \vert^{d'_2}-1$.  The next $2 \vert x \vert^{d'_2}$ blocks contain the necessary information when $i = 1$ and $i'$ again ranges from 0 to $\vert x \vert^{d'_2}-1$.  (These groups contain $2 \vert x \vert^{d'_2}$ blocks rather than just $\vert x \vert^{d'_2}$ blocks because we need two blocks from $w$ and two from one of the $w'_k$s, but each block of $y$ contains only half this information.)  More generally, if we let $\alpha$ range from 0 to $\vert x \vert^{d_2}-1$ (the range of $i$) and we let $\beta$ be the even numbers in the range 0 to $\vert x \vert^{d'_2}-1$ (the range of $i'$), then the $(2 \alpha \vert x \vert^{d'_2} + \beta)$th block of $y$ is $\langle (w)_{f_1(\alpha)}, (w'_{\alpha})_{f'_1(\beta)} \rangle$, and the $(2 \alpha \vert x \vert^{d'_2} + \beta + 1)$th block of $y$ is $\langle (w)_{f_2(\alpha)}, (w'_{\alpha})_{f'_2(\beta)} \rangle$.

Due to our selection of $e_2$ above, we can treat each value of the index $j$ as being of the form $\langle i, i' \rangle$, ignoring any extra bits that result from the fact that $\log(\mbox{max}(j)) \geq \log(\mbox{max}(ii'))$.  Then we define $g_1(j) = g_1(\langle i, i' \rangle) = 2i\vert x \vert^{d'_2} + 2i'$, and we define $g_2(j) = g_1(j) + 1$.  And finally, by viewing the index $j$ and the blocks of $y$ as we have defined them above, the predicate $S(x, j, (y)_{g_1(j)}, (y)_{g_2(j)})$ is equal to $R(x, i, (w)_{f_1(i)}, (w)_{f_2(i)}, i', (w'_i)_{f'_1(i')}, (w'_i)_{f'_2(i')})$.
\end{proof}

\begin{cor}
\label{xiLINH-upper-bound}
For all $k \geq 0$, $(\xi)^{\lin}_k = (\xi)^{\lin}_1 = \xi\DLIN \subseteq \NP$.
\end{cor}
\begin{proof}
The first equality follows from Theorem~\ref{xi-collapse}, and the second equality from the definition of $(\xi)^{\lin}_1$.  The inclusion $\xi\DLIN \subseteq \NP$ follows from the fact that an $\NP$ machine could existentially guess the string $w$, and then deterministically try each of the polynomially-many values of $i$ to decide a language in $\xi\DLIN$.
\end{proof}

We now use the following result from~\cite{gupta96}, which was largely based on work originally published in~\cite{ppst83}.

\begin{thm}
\label{dlin-upper-bound}
$\DLIN \subseteq \Sigma_2$-$\TISP[n,\frac{n}{\log^*n}]$.
\end{thm}

The result actually proved was that $\DTIME[t(n)] \subseteq \Sigma_2[\frac{t(n)}{\log^*n}]$ for all time-constructible $t(n) \geq n\log^*n$, but as the lower bound on $t(n)$ is not relevant to the space bound on the simulating machine in the proof, the same proof can be used to show Theorem~\ref{dlin-upper-bound}.

\begin{cor}
$\EuLINH \subseteq \xi\Sigma_2$-$\TISP[n,\frac{n}{\log^*n}]$.
\end{cor}
\begin{proof}
From Theorem~\ref{dlin-upper-bound}, we have that $(\xi)^{\lin}_1 \subseteq \xi\Sigma_2$-$\TISP[n,\frac{n}{\log^*n}]$.  From Corollary~\ref{xiLINH-upper-bound}, we then have that $(\xi)^{\lin}_k \subseteq \xi\Sigma_2$-$\TISP[n,\frac{n}{\log^*n}]$ for all $k \geq 0$.  Finally, since every language $L \in \EuLINH$ must be in some precise level of $\EuLINH$, and thus in some precise level of $\xiLINH$, the result follows.
\end{proof}

\section{Conclusion}
We believe we have shown that the hierarchy $\EuLINH$ is an interesting hierarchy closely connected with the $\NLOGSPACE =\NP$ question. This is evidenced both by the containments 
$$\NTISP[n^{k(1-\epsilon)}, n^\epsilon)]\subseteq(\EXIST\SUNIV)^{\lin}_{k}\subseteq  \NTISP[n^{k+1}, n] $$ and by the fact that if our hierarchy collapses then $\NLOGSPACE \neq \NP$. We would like briefly to conclude with some future directions for research. One immediate question is whether the containments $\NTISP[n^{k(1-\epsilon)}, n^\epsilon]\subseteq(\EXIST\SUNIV)^{\lin}_{k}\subseteq  \NTISP[n^{k+1}, n] $ can be made tighter?  We have given some evidence for this in the preceding section, but is an exact relationship between our hierarchy and a nondeterministic time space class possible? It would also be interesting to study what happens when one considers the hierarchies like $\EuLINH$ where one has nonconstantly many alternations. Given that the total size of the string that would be fed into the machine in this case could be nonlinear in the input size, it would probably be worthwhile to consider quasi-linear variants of our class as well. 


\bibliographystyle{plain}

\begin{thebibliography}{1}

\bibitem{bbg98}
S.~Bloch, J.~Buss and J.~Goldsmith.
\newblock Sharply bounded alternation and quasilinear time. 
\newblock Theory of Computing Systems, 31(2):187-214, March 1998.

\bibitem{cs76}
A.~K.~Chandra and L.~J.~Stockmeyer.
\newblock Alternation. 
\newblock IEEE Symposium on Foundations of Computer Science, pp\@. 98--106, 1976.

\bibitem{fortnow00}
L.~Fortnow.
\newblock Time-space tradeoffs for satisfiability. 
\newblock Journal of Computer and System Sciences, 60(2):337-353, April 2000. 

\bibitem{flmv06}
L.~Fortnow and R.~Lipton and D.~van~Melkebeek and A.~Viglas. 
\newblock Time-space lower bounds for satisfiability. 
\newblock Journal of the ACM, 52(6):835-865, November 2005.

\bibitem{gupta96}
S.~Gupta.
\newblock Alternating Time Versus Deterministic Time: A Separation.
\newblock {\em Mathematical Systems Theory}, pp\@. 661-672, 1996.

\bibitem{kadin88}
Jim Kadin.
\newblock The Polynomial Time Hierarchy Collapses if the Boolean Hierarchy Collapses. 
\newblock SIAM Journal on Computing, 17(6): 1263--1282 (1988).

\bibitem{hs65}
J.~Hartmanis and R.~Stearns.
\newblock On the computational complexity of algorithms. 
\newblock Transactions of the American Mathematical Society, Vol\@. 117, pp\@. 285--306\@, 1965. 

\bibitem{hp}
P.~H\'{a}jek and P.~Pudl\'{a}k.
\newblock {\em Metamathematics of First-Order Arithmetics}.
\newblock Springer-Verlag, 1993.

\bibitem{immerman88}
N.~Immerman.
\newblock Nondeterministic Space is Closed Under Complement.
\newblock SIAM Journal on Computing, Vol\@. 17, pp\@. 935--938, 1988. 

\bibitem{johnson90}
D.~S.~Johnson.
\newblock A Catalog of Complexity Classes.
\newblock In {\em The Handbook of Theoretical Computer Science}, Volume A. J. Van Leeuwen, Ed\@.,
pp\@. 68--161, MIT Press, 1990.

\bibitem{lv99}
R.~Lipton and A.~Viglas
\newblock On the Complexity of SAT.
\newblock IEEE Symposium on Foundations of Computer Science, pp\@. 459--464. 1999.

\bibitem{n70}
V.A.~\Nepom.
\newblock Rudimentary predicates and Turing computations.
\newblock {\em Dokl. Acad. Nauk}, Vol\@. 195, pp\@. 282--284, 1970. transl. 
Vol. 11, pp\@. 1462--1465, 1970.

\bibitem{ppst83}
W.~J.~Paul, N.~Pippenger, E.~Szemeredi and W.~T.~Trotter.
\newblock On determinism versus non-determinism and related problems.
\newblock In {\em Proceedings of the 24th Annual IEEE Symposium on Foundations of Computer Science}, pp\@. 429--438, 1983.

\bibitem{sfm78}
J.~Seiferas and M.~Fischer and A.~Meyer. 
\newblock Separating nondeterministic time complexity  classes. 
\newblock Journal of the ACM, 25:146Ð167, 1978.

\bibitem{sz87}
R.~Szelepcsenyi.
\newblock The method of forcing for nondeterministic automata.
\newblock Bulletin EATCS 33, pp\@. 96--100, 1987. 

\bibitem{williams2005}
\newblock Ryan Williams: 
\newblock Better Time-Space Lower Bounds for SAT and Related Problems. 
\newblock IEEE Conference on Computational Complexity (CCC 2005), pp\@. 40--49, 2005.

\bibitem{zak83}
S.~\Zak. 
\newblock A Turing machine time hierarchy. 
\newblock Theoretical Computer Science, 26:327Ð333, 1983. 

\end{thebibliography}

\end{document}